\documentclass{revtex4}
\usepackage{amsmath}
\usepackage{amssymb}
\usepackage{graphicx}

\begin{document}

\title{Creation/annihilation of wormholes \\supported by the Sine-Gordon phantom (ghost) field}

\author{
Vladimir Dzhunushaliev,$^{1,2,3}$
\footnote{
Email: vdzhunus@krsu.edu.kg}
Vladimir Folomeev,$^{2,3}$
\footnote{Email: vfolomeev@mail.ru}
}
\affiliation{$^1$Department of Physics and Microelectronic
Engineering, Kyrgyz-Russian Slavic University, Bishkek, Kievskaya Str.
44, 720021, Kyrgyz Republic \\ 
$^2$Institute of Physicotechnical Problems and Material Science of the NAS
of the
Kyrgyz Republic, 265 a, Chui Street, Bishkek, 720071,  Kyrgyz Republic \\
$^3$Institut f\"ur Physik, Universit\"at Oldenburg, Postfach 2503
D-26111 Oldenburg, Germany
}

\begin{abstract}
The possible process of creation/annihilation of traversable wormholes in the model with phantom (ghost) scalar
field is described. It is shown that such process can be realized only for some special choice of a potential energy,
in particular, for the Sine-Gordon potential.
\end{abstract}
\maketitle

\section{Introduction}

In the paper \cite{Morris:1988cz}, the \emph{conditions} of existence of  traversable wormholes have been considered.
In summary, the following conclusion had been done: \emph{"\ldots However, any hope that they (wormholes) might be constructable must rely on the future discovery of an exotic field or quantum state of known fields with tension that exceeds energy density on macroscopic length scales \ldots"}.   
The last condition means violation of the weak energy condition (WEC) which
 states that $\rho+p\geq 0$, where $\rho$ and $p$ are effective energy density and pressure of matter.
It can be done in a few ways (for a review, see \cite{Visser}).
One of possible ways of violation of the WEC is using  some phantom (or ghost) scalar fields as a source of matter.
There are some works in this direction
 (see, e.g., Refs.~\cite{Sushkov, Lobo}). In these papers some effective
hydrodynamical energy-momentum tensor with the violated WEC was chosen as a source of matter. But a distribution of this
matter was added by hand and, accordingly, the 
nonself-consistent models of the traversable wormholes were
considered. In Ref.~\cite{Dzhunushaliev:2007cs} the self-consistent model of the traversable wormhole supported by
two interacting phantom and ghost scalar fields was considered.

But these investigations just show that there is a possibility of existence of static wormhole-like solutions, leaving aside the question
of dynamical evolution of the models. Of course, investigation of such dynamical problem demands a consideration of a
self-consistent problem about evolution of a  traversable wormhole with time. In this paper we consider more  modest problem.
We try to show that there exists a set of static wormhole-like solutions with smoothly varying throat radius up to the zero value.
It might be interpreted as
 a \emph{possible} process of creation/annihilation of wormholes by adding/subtraction of phantom (ghost) matter to
some spherically symmetric object created by phantom (or ghost) matter with the Sine-Gordon potential energy. Our main purpose is to show that
for such  phantom field there exists a sequence of wormhole solutions with a vanishing throat radius.  This might be interpreted as
follows: increasing or decreasing the phantom matter, one can decrease or increase the throat radius.
If to change a phantom  mass of the throat in such a way to get a zero throat radius, then, as a result, one will obtain two spherically symmetric
solutions connected (strictly speaking)  in one point.
If there is a possibility to break up these two spaces in this point, then one might obtain two separate spaces, i.e., the wormhole will be broken up.
One can suppose that the inverse process of creation of a wormhole also exists.

\section{Wormholes supported by a phantom Sine-Gordon scalar field}

Let us consider a gravitating system with one phantom scalar field $\varphi$ with the Lagrangian
\begin{equation}
\label{lagrangian}
  L =-\frac{R}{16\pi G}-
      \frac{1}{2}\partial_\mu \varphi \partial^\mu
        \varphi -V(\varphi)~,
\end{equation}
where $R$ is the scalar curvature, $G$ is the Newton's gravitational constant, 
and $V$ is the  Sine-Gordon potential with the reversed sign
\begin{equation}
\label{pot_mex2}
	V=\frac{m^4}{\lambda} \left[
		\cos\left(\frac{\sqrt{\lambda}}{m}\varphi\right)-1
	\right].
\end{equation}
Here $m$ is a mass of the field, and $\lambda$ is a coupling constant. The corresponding energy-momentum tensor will then be:
\begin{equation}
\label{emt}
    T^k_i=
    -\partial_i \varphi \partial^k \varphi-
        \delta^k_i \left[
            -\frac{1}{2}\partial_\mu \varphi \partial^\mu
            \varphi-V(\varphi)
        \right]~,
\end{equation}
and variation of the Lagrangian \eqref{lagrangian} gives the gravitational and field equations in the form
\begin{eqnarray}
\label{Einstein-gen}
    G_{i}^k &=&  8\pi G T^k_i,
\\
\label{field-gen}
	\frac{1}{\sqrt{-g}}\frac{\partial}{\partial 	x^\mu}
	\left(
		\sqrt{-g}\,\, g^{\mu\nu} \frac{\partial \varphi}{\partial x^\nu}
	\right) &=& \frac{\partial V}{\partial
	\varphi}.
\end{eqnarray}
Let us take the spherically symmetric metric in the form
\begin{equation}
\label{metric_wh}
ds^2=e^{2 F(r)}dt^2-\frac{dr^2}{A(r)}-(r^2+r_0^2)(d\theta^2+\sin^2\theta d\phi^2),
\end{equation}
where the metric functions $F(r)$ and $A(r)$ depend only on the radial coordinate $r$.

\begin{figure}[ht]
\begin{center}
  \includegraphics[width=13cm]{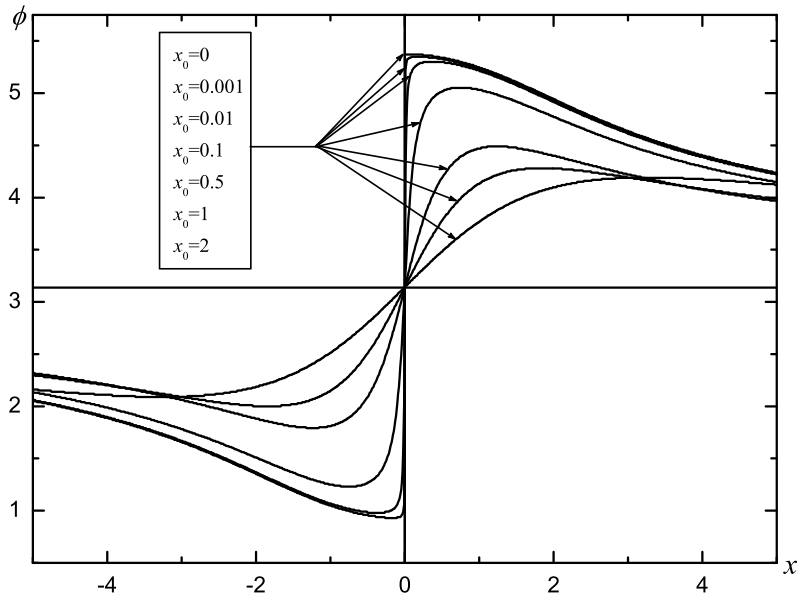}
\vspace{-1cm}
 \caption{The behavior of the scalar field $\phi$ depending on the throat radius $x_0$. $\beta=1$ for all lines.
 Asymptotically $\phi\rightarrow \pi$.}
\label{phi_wh_sine}
\end{center}
\end{figure}

Introducing  new dimensionless variables $\phi=(\sqrt{\lambda}/m)\varphi,\, x=m r$, one can obtain from \eqref{emt}, \eqref{Einstein-gen} and \eqref{field-gen}
the following equations
\begin{eqnarray}
\label{ein_wh_sine}
	-\frac{A^\prime}{A}x+\frac{1}{A}+\frac{x^2}{x^2+x_0^2}-2&=&
	\frac{x^2+x_0^2}{A}\beta\left(-\frac{A}{2}\phi^{\prime 2}
	+\cos\phi-1\right),
\\
	\phi^{\prime\prime}+\left(-\frac{\beta}{2}\frac{x^2+x_0^2}{x}\phi^{\prime 2}
	+\frac{A^\prime}{A}+\frac{1}{x}+\frac{x}{x^2+x_0^2}\right)\phi^\prime&=&
	\frac{1}{A}\sin\phi,
\label{field_wh_sine}
\end{eqnarray}
where $\beta=m^2/\lambda$. The boundary conditions follow from the equations above. In the wormhole case, one has
\begin{equation}
\label{ini_wh_sine}
	A(0)=1+2 \beta x_0^2, \quad \phi(0)=\pi, \quad
	\phi^\prime(0)=\sqrt{\frac{2}{\beta x_0^2}},
\end{equation}
and in the spherically symmetric case
\begin{equation}
\label{ini_sphera_sine}
	A(0)=1, \quad \phi(0)=const, \quad \phi^\prime(0)=0.
\end{equation}
Solving  numerically equations \eqref{ein_wh_sine} and \eqref{field_wh_sine} with the boundary conditions \eqref{ini_wh_sine} and \eqref{ini_sphera_sine} at different values of $x_0$ and with some $\beta=1$, one can obtain the results presented in Fig. \ref{phi_wh_sine}. 
One can see that asymptotically (at $x\rightarrow \pm \infty$) $\phi\rightarrow \pi$.

That is, as it was pointed out in Introduction, we have obtained such a sequence of solutions with $x_0 \rightarrow 0$ that
the solution at $x_0 = 0$ also exists. The last means that the spherically symmetric non-wormhole solution has been found. This limit solution
corresponds to two different spaces glued in one point. If we divide them in this point then we will obtain two separate spaces. Both these spaces
contain spherically symmetric solutions of the corresponding equations \eqref{emt}-\eqref{field-gen}. 
From the physical point of view, it is clear that solutions with different $x_0$ differ from each other by different amount of the phantom matter in the wormhole.
That is why our interpretation of this result consists in assumption that the increase/decrease of phantom matter decreases/increases
the throat radius that results in break up of the throat and appearance of two separate spaces. One can assume that the inverse process,
leading to creation of a wormhole, also exists.

Note that all wormhole solutions obtained above are not asymptotically flat (the spacetime is anti-de Sitter one).
One can try to find a similar sequence
of solutions with an asymptotically flat spacetime. It will be done in the next section with use of the Mexican hat potential.
Let us note right now that such a sequence of asymptotically flat solutions does not exist.

\section{Wormholes supported by a phantom Mexican hat}

In this section we follow to Ref.~\cite{Kodama:1978dw}. Let us consider a model with the Mexican hat potential
\begin{equation}
\label{pot_mex}
V=\frac{1}{2}\left(\frac{\mu}{f}\right)^2\left[\left(1-f^2\varphi^2\right)^2-1\right].
\end{equation}
Here $\mu$ and $f$ are some constants.
Using \eqref{emt}, \eqref{Einstein-gen} and \eqref{field-gen},
one can obtain the $(_t^t)$ and $\left[(_t^t)-(_x^x)\right]$ components of the Einstein equations~\eqref{Einstein-gen},  respectively
\begin{eqnarray}
\label{ein_wh_1}
	-\frac{A^\prime}{A}x+\frac{1}{A}+\frac{x^2}{x^2+x_0^2}-2&=&
	\frac{x^2+x_0^2}{f^2 A}\left\{
		-\frac{A}{2}\phi^{\prime 2}
		+\frac{1}{2}\left[\left(1-\phi^2\right)^2-1\right]
	\right\},
\\
	\frac{2x^2}{x^2+x_0^2}-2-\frac{A^\prime}{A}x+
	2x F^\prime&=&-\frac{x^2+x_0^2}{f^2}\phi^{\prime 2},
	\label{ein_wh_2}
\end{eqnarray}
and the scalar field equation
\begin{equation}
\label{field_wh_1}
\phi^{\prime\prime}+\left(F^\prime+\frac{2x}{x^2+x_0^2}+\frac{1}{2}\frac{A^\prime}{A}\right)\phi^\prime=\frac{2}{A}\phi\left(1-\phi^2\right),
\end{equation}
where a prime denotes differentiation with respect to $x$. Here and further we will use the
dimensionless variables $x=\mu r, \phi=f \varphi$, and $8\pi G=c=1$.

Using Eq. \eqref{ein_wh_2}, one can rewrite Eq. \eqref{field_wh_1} as follows
\begin{equation}
\label{field_wh_2}
\phi^{\prime\prime}+\left(-\frac{1}{2f^2}\frac{x^2+x_0^2}{x}\phi^{\prime 2}
+\frac{A^\prime}{A}+\frac{1}{x}+\frac{x}{x^2+x_0^2}\right)\phi^\prime=\frac{2}{A}\phi\left(1-\phi^2\right).
\end{equation}

We solve the system of equations~\eqref{ein_wh_1} and \eqref{field_wh_2} with the following boundary conditions at $x=0$:
\begin{equation}
\label{ini1_wh}
    \phi(0)=const,      \quad \phi^\prime(0)=\sqrt{2}\frac{f}{x_0}, \quad
    A(0)            =1-\frac{x_0^2}{2 f^2}\left[\left(1-\phi(0)^2\right)^2-1\right].
\end{equation}
If one wants to start with $A(0)=1$ and to get anti-symmetric solutions with respect to $x=0$, it is necessary to take $\phi(0)=0$.

\begin{figure}[ht]
\begin{center}
  \includegraphics[width=13cm]{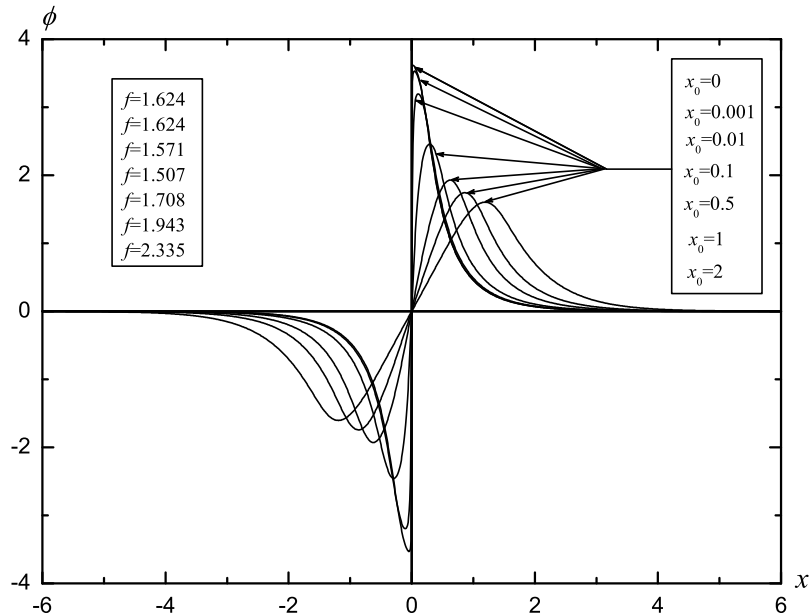}
\vspace{-1cm}
 \caption{The behavior of the scalar field $\phi$ depending on the throat radius $x_0$. On the left, the
 corresponding values of the parameter $f$ from the potential \eqref{pot_mex} are presented.}
\label{phi_wh}
\end{center}
\end{figure}

The obtained results are presented in Fig. \ref{phi_wh}. As one can see, the regular wormhole ($x_0\neq 0$) and spherically symmetric ($x_0= 0$) solutions only exist at different values of the parameter $f$ from the potential \eqref{pot_mex}. Every solution for a given $x_0$ exists only at definite $f$. In this sense, $f$ is an eigenvalue of the problem.
This means that the sequence of solutions necessary for us does not exist since for each $f$ only one solution with a special value of $x_0$ does exist.

Summarizing,  we have shown that there exists some engineering possibility to create a sequence of wormholes with
the vanishing throat radius in the asymptotically AdS spacetime.
This means that it is possible to break up the wormhole into two separate  spherically symmetric solutions
by changing the phantom mass of the wormhole.
It seems to be physically sensible that the inverse process is also possible: by changing mass in two regions of space containing spherically
symmetric distribution of phantom matter created by scalar field with the Sine-Gordon potential, it is possible to create a wormhole
between these two regions.

\section*{Acknowledgements}
V.D. is grateful to the Research Group Linkage Programme of the Alexander von Humboldt Foundation for the support of this research. V.F. would like to thank the German Academic Exchange Service  (DAAD) for financial support. We  would like to express our gratitude to the Department of Physics of the Carl von Ossietzky University of Oldenburg  and, specially, to Prof. J. Kunz and Prof. B. Kleihaus for fruitful discussions.

\end{document}